# Software Vulnerabilities, Banking Threats, Botnets and Malware Self-Protection Technologies

Wajeb Gharibi[1], Abdulrahman Mirza[2]

[1] **Computer Networks Department, Computer Science & Information Systems College, Jazan University
Jazan 82822-6694, Saudi Arabia**

[2] **Information Systems Department, King Saud University, Center of Excellence in Information Assurance
Riyadh 11482, Saudi Arabia**

### Abstract

Information security is the protection of information from a wide range of threats in order to ensure success business continuity by minimizing risks and maximizing the return of investments and business opportunities.
In this paper, we study and discuss the software vulnerabilities, banking threats, botnets and propose the malware self-protection technologies.

## 1. Introduction

Nowadays, there is a huge variety of cyber threats that can be quite dangerous not only for big companies but also for an ordinary user, who can be a potential victim for cybercriminals when using unsafe system for entering confidential data, such as login, password, credit card numbers, etc. Among popular computer threats it is possible to distinguish several types depending on the means and ways they are realized. They are: malicious software (malware), DDoS attacks (Distributed Denial-of-Service), phishing, banking, exploiting vulnerabilities, botnets, threats for mobile phones, IP-communication threats, social networking threats and even spam. All of these threats try to violate one of the following criteria: confidentiality, integrity and accessibility.

Obviously that hackers use the malicious programs to gain control of targeted computer in order to use it further for other types of cyber attacks. As a result, malicious software has turned into big business and cyber criminals became profitable organizations and able to perform any type of attack. An understanding of today's cyber threats is a vital part for safe computing and ability to counteract the cyber invaders.

Our paper is organized as follows: Section 2 demonstrates the software vulnerabilities. Section 3 proposes banking threats. Section 4 defines botnets. Conclusions have been made in Section 5.

## 2. Software Vulnerabilities

The term 'vulnerability' is often mentioned in connection with computer security, in many different contexts. It is associated with some violation of a security policy. This may be due to weak security rules, or it may be that there is a problem within the software itself. In theory, all computer systems have vulnerabilities [1-5].

MITRE, a US federally funded research and development group, focuses on analyzing and solving critical security issues. The group has defined the followings:

Definition 2.1 A universal vulnerability is a state in a computing system (or set of systems) which either allows an attacker to execute commands as another user, or to access data that is contrary to the specified access restrictions, or to pose as another entity to conduct a denial of service.

Definition 2.2 An exposure is a state in a computing system (or set of systems) which is not a universal vulnerability, but either allows an attacker to conduct information gathering activities or hide activities or includes a capability that behaves as expected, but can be easily compromised.

It is a primary point of entry that an attacker may attempt to use to gain access to the system or data is considered a problem according to some reasonable security policy.

Microsoft Windows, the operating system most commonly used on systems connected to the Internet, contains multiple, severe vulnerabilities. The most commonly exploited are in IIS, MS-SQL,

Internet Explorer, the file serving and message processing services of the operating system itself [6, 7].

A vulnerability in IIS, detailed in Microsoft Security Bulletin MS01-033, is one of the most exploited Windows vulnerabilities ever. A large number of network worms have been written over the years to exploit this vulnerability, including 'CodeRed' which was first detected on July 17th 2001 and is believed to have infected over 300000 targets. Still some versions of CodeRed worm are spreading throughout the Internet [8].

Spida Network Worm, detected almost a year after CodeRed appeared, relied on an exposure in MS-SQL server software package to spread.

Slammer Network Worm, detected in late January 2003, used an even more direct method to infect Windows systems running MS-SQL server: a buffer overflow vunerability in one of the UDP packet handling subroutines. As it was relatively small - 376 bytes - and used UDP, a communication protocol designed for the quick transmission of data, Slammer spread at an almost incredible rate. Some estimate the time taken for Slammer to spread across the world at as low as 15 minutes, infecting around 75000 hosts [9].

However, Lovesan Worm, detected on 11th August 2003, used a much more severe buffer overflow in a core component of Windows itself to spread. This vulnerability is detailed in Microsoft Security Bulletin MS03-026.

Sasser Worm was first appeared at the beginning of May 2003, exploited another core component vulnerability, this time in the Local Security Authority Subsystem Service (LSASS). Sasser spread rapidly and infected millions of computers world-wide, at an enormous cost to business [10].

From last incidents also it is possible to note that epidemic of Worm Kido/Conficker/Downadup which as one of distribution methods used vulnerability MS08-067 in service "Server" (http://www.microsoft.com/technet/security/Bulletin/MS08-067.mspx).

Inevitably, all operating systems contain vulnerabilities and exposures which can be targeted by hackers and virus writers. Although Windows vulnerabilities receive the most publicity due to the number of machines running Windows, Unix and MacOS have also their own weak spots.

## 3. Banking Threats

Definition 3.1 Banking - one of the remote bank service kinds at which management is made through the Internet.

In 2007, antivirus vendors saw a huge increase in the number of malicious programs targeting banks (financial malware) according to Kaspersky Lab stats (Figure 1). In spite of the lack of clear information from the financial sector, this indicates a corresponding increase in the number of banks attacks.

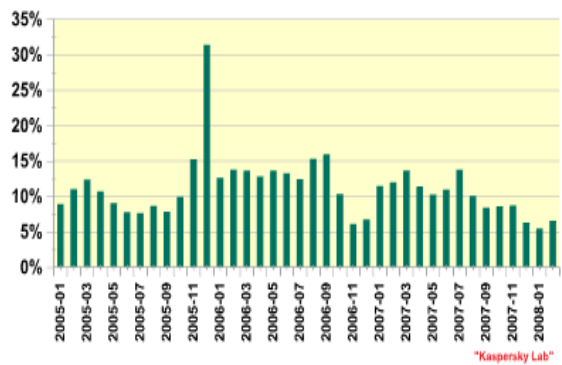

Figure1. Percentage of financial malware among all malicious programs detected

Notwithstanding an increased number of attacks, as the graph above shows, the percentage of financial malware detected each month is dropping. The reasons for this are detailed below:

- Malware authors constantly change their programs in order to evade detection by antivirus solutions. However, if the changes made are minor, AV vendors will still be able to detect new malware samples using signatures created for previous variants.

- The graph above covers only financial malware. However, banking attacks are usually a multi-step process: social engineering, phishing, and the use of Trojan-Downloaders which then download the financial malware. It's easier for the criminals to modify the Trojan-Downloader programs (which are usually smaller in size, and generally less complex) than the financial malware itself.

In 2007, there was an upsurge in the number of password stealing Trojans designed to steal all data entered into web forms. These target the most popular browsers i.e. Internet Explorer, Opera and Firefox. Such Trojans can obviously be used to steal credit cards, and using such malware may be enough to breach a bank's defenses – it all depends on the sophistication of the security measures employed [11, 12].

Actually, malicious programs delivered via email are more likely to attract the attention of antivirus vendors and financial institutions, not to mention the media and end users. Stealth is the key factor in the success of attacks on financial institutions, so conducting a drive-by download using exploits is obviously an attractive method. Moreover, it is a significant factor in terms of evading quick

238

detection by antivirus solutions – malicious programs which infect victim systems via the web are hosted on a web server. This means that the cyber criminals using these programs to conduct attacks can modify the malicious files very easily using automated tools – a method known as server-side polymorphism. In contrast to regular host polymorphism (where the algorithm used to modify the code is contained in the body of the malicious program) it's impossible for antivirus researchers to analyze the algorithm used to modify the malware, as it's located on the remote server.

In addition, some of the more sophisticated Trojan-Downloaders used to deliver financial malware to its eventual destination are designed to self-destruct (or 'melt') once they have successfully or unsuccessfully downloaded the financial malware.

The use of Transaction Authorization Numbers (TAN) for signing transactions makes gaining access to accounts somewhat more complex. The TAN may come from a physical list issued to the account holder by the financial organization or it may be sent via SMS.

Another method used by cyber criminals is to redirect traffic. There are several ways of doing this, and the easiest of these is to modify the Windows "hosts" file which is located in the %windows%\system32\drivers\etc directory, can be used to bypass DNS (Domain Name Server) lookups. DNS is used to translate domain names, such as www.kaspersky.com, into an IP address. Domain names are used purely for convenience; it's the IP addresses which are used by computers. If the host files are modified to point a specific domain name to the IP address of a fake site, the computer will be directed to that site.

Another method for redirecting traffic is to modify the DNS server settings. Instead of trying to bypass DNS lookups, the settings are changed in such a way that the machine uses a different, malicious, DNS server for the lookups. Most people surfing from home use the DNS server belonging to their ISP for lookups. As a result, the vast majority of this type of attack has been directed at workstations. However, when a router is used to access the internet, by default it's the router performing DNS lookups and passing them on to the workstations.

Yet another method which can be used to redirect traffic is to place a Trojan on the victim machine which monitors the sites visited. As soon as the user connects to a banking site (or that of another financial organization) the Trojan will redirect the traffic to a fake website. The traffic may be redirected from an HTTPS site to an HTTP (potentially insecure) site. In such cases, the Trojan is usually able to suppress any warning message issued by the browser.

## 4. BOTNETS

Botnets have been in existence for about 10 years; experts have been warning the public about the threat posed by botnets for more or less the same period.

Definition 4.1 A botnet is a network of computers made up of machines infected with a malicious backdoor program. The backdoor enables cybercriminals to remotely control the infected computers (which may mean controlling an individual machine, some of the computers making up the network or the entire network).

Malicious backdoor programs that are specifically designed for the use of creating botnets are called bots. Botnets have vast computing power. They are used as a powerful cyber weapon and are an effective tool for making money illegally. The owner of a botnet can control the computers which form the network from anywhere in the world – from another city, country or even another continent. Importantly, the Internet is structured in such a way that a botnet can be controlled anonymously [13].

Botnets can be used by cybercriminals to conduct a wide range of criminal activities, from sending spam to attacking government networks:

- Sending spam - the most common use for botnets (over 80% of spam is sent from zombie computers).

- DDoS atacks - using tens or even hundreds of thousands of computers to conduct DDoS (Distributed Denial of Service) attacks.

- Anonymous Internet access; cybercriminals can access web servers using zombie machines and commit cybercrimes such as hacking websites or transferring stolen money.

- Selling and leasing botnets. One option for making money illegally using botnets is based on leasing them or selling entire networks. Creating botnets for sale is also a lucrative criminal business.

- Phishing; a botnet allows phishers to change the addresses of phishing pages frequently, using infected computers as proxy servers. This helps conceal the real address of the phishers' web server.

- Theft of confidential data; botnets help to increase the haul of passwords (passwords to email and ICQ accounts, FTP resources, web services etc.)

- There are currently only two known types of

botnet architecture:

a) Centralized botnets; in this type of botnet, all computers are connected to a single command-and-control center or C&C. The C&C waits for new bots to connect, registers them in its database, tracks their status and sends them commands selected by the botnet owner from a list of bot commands. All zombie computers in the botnet are visible to the C&C. The zombie network owner needs access to the command and control center to be able to manage the centralized botnet.

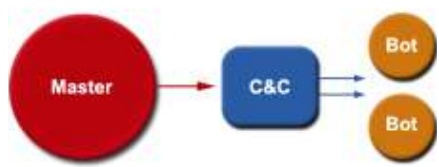

Figure 2 Centralized topology (C&C)

Centralized botnets are the most widespread type of zombie network. Such botnets are easier to create, easier to manage and they respond to commands faster. However, it is also easier to combat centralized botnets, since the entire zombie network is neutralized if the C&C is put out of commission.

b) Decentralized or P2P (peer-to-peer) botnets; in a decentralized botnet, bots connect to several infected machines on a bot network rather than to a command and control center. Commands are transferred from bot to bot: each bot has a list of several 'neighbors', and any command received by a bot from one of its neighbors will be sent on to the others, further distributing it across the zombie network. In this case, a cybercriminal needs to have access to at least one computer on the zombie network to be able to control the entire botnet.

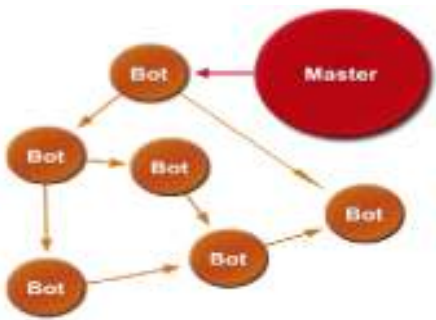

Figure 3 Decentralized topology (P2P)

In practice, building decentralized botnets is not easy task, since each newly infected computer needs to be provided with a list of bots to which it will connect on the zombie network.

Today, botnets are among the main sources of illegal income on the Internet and they are powerful weapons in the hands of cybercriminals [14].

What makes botnets increasingly dangerous is that they are becoming easier and easier to use. In the near future, even children will be able to manage them. The ability to gain access to a network of infected computers is determined by the amount of money cybercriminals have at their disposal rather than whether they have specialized knowledge. Additionally, the prices in the well-developed and structured botnet market are relatively low.

## 5. Malware Self-Protection Technologies

Trying to hide the presence of a malicious component in binary code of the program or in script of the web-page, hackers use various techniques, such as: encryption, or polymorphism, or obfuscation, or packing. Thus the malicious program complicates the process of signature detection of the code by the anti-virus scanner. Such methods of protection have received the name passive.

*Encryption* is the universal mechanism which can be applied for protection of code as well, as for ciphering the data of the user and demanding the payment for their decoding, as it has been realized in one of the first viruses - Cascade which contained polymorphic encryptor and ciphered each new copy of a virus a unique key, and as dangerous malicious functional, for example, as it was implemented with virus GpCode [15].

Definition 5.1 Polymorphism — a technology that allows a self-replicating program to fully or partially modify its outward appearance and/or the structure of its code during the replication process.

Definition 5.2 Obfuscation — a combination of approaches used to obscure the source code of a program.

This is designed to make the code as difficult as possible to read and analyze it while retaining full functionality. Obfuscation technologies can be applied at the level of any programming language (including high level, script and assembler languages). Examples of very simple obfuscation include adding neutral instructions (which do not alter program functionality) to the code or making the code harder to read by using an excessive number of unconditional skips (or unconditional changeovers disguised as conditional skips).

There are many approaches that can be applied for these purposes (dynamic code generators,

polymorphism, etc.), but in most cases, the authors of malicious programs don't spend much time or effort on developing these types of mechanisms. They use a much simpler solution in order to achieve their goals: so-called packers. These are utilities that use dedicated algorithms to encode the target executable program while retaining its functionality. The use of packers makes a malicious user's task much easier: in order to prevent an antivirus program from detecting an already known malicious program, the author no longer has to rewrite it from scratch - all he has to do is re-pack it with a packer that is not known to the antivirus program. The result is the same, and the costs are much lower.

The continued acceleration of the increase in new malicious programs (Figure 4) is now accompanied by an increase in the total number of malicious programs which actively combat antivirus solutions. First and foremost, this involves virus writers using rootkit technologies in order to increase the lifespan of a virus in the infected system: if a malicious program has stealth capabilities, then it is less likely to end up in an antivirus company's database [16, 17].

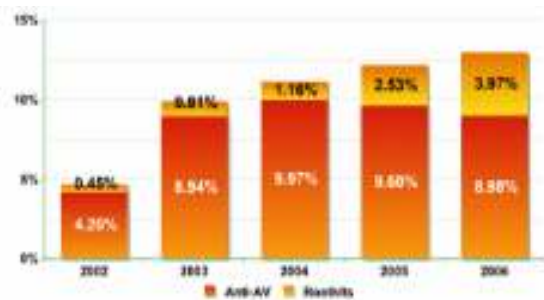

Figure 4 Increase in the number of new modifications of malicious programs that actively combat security solutions.
Source: Kaspersky Lab

There have always been malicious programs that have actively defended themselves. Self-defense mechanisms include:

- Performing a targeted search of the system for an antivirus product, firewall or other security utility, followed by disrupting the functioning of that utility. An example might be a malicious program that searches for a specific antivirus product in the process list and subsequently attempts to disrupt the functioning of that antivirus.
- Blocking files and opening them with exclusive access as a counter measure against file scanning by the antivirus.
- Modifying the hosts file in order to block access to antivirus update sites.
- Detecting query messages sent by the security system (for example, a firewall window with an inquiry such as "Allow this connection?") and imitating a click on the "Allow" button.

Rootkit, or as it is usually referred to, bootkit because it runs during the boot sequence, is based on the eEye Bootroot code. Essentially, it's not so much a separate piece of malware as a tool to hide Trojans…any Trojans. Consequently, it seems a reasonable conclusion that Sinowal is being shared (possibly for a fee) in certain circles and that we haven't seen the last of it by any means [18-21].

## 6. Conclusions

The most common types of contemporary threats are considered as well, as mechanism of malware self-protection aimed to counteract against antiviruses. New areas of hackers' attacks were highlighted in this paper. Thus, it gives us clear understandings what is going on in the world of cyber security and helps to protect our computer systems from undesirable intruders and confidential data theft.

Nowadays, we use more and more online services in Internet which can be threat of personal information stealing by third party. It is getting more important to keep our data in secure place protected by antivirus and DLP (Data Leakage Protection) systems.